\documentclass[twocolumn,showpacs]{revtex4}

\usepackage{graphicx}
\usepackage{bm}
\usepackage{amssymb}
\usepackage{amsmath}

\newcommand{\Z}{\mathbf{Z}}
\newcommand{\F}{\mathbf{F}}
\newcommand{\N}{\mathbf{N}}
\newcommand{\C}{\mathbf{C}}
\newcommand{\R}{\mathbf{R}}
\newcommand{\onemat}[0]{{\mathbf 1}}

\newtheorem{theorem}{Theorem}

\def\ket#1{\left|#1\right>}
\def\bra#1{\left<#1\right|}

\begin{document}

\title{Efficient Decoupling Schemes Based on Hamilton Cycles}

\author{Martin R{\"o}tteler}
\email{mroetteler@iqc.ca}
\affiliation{
Institute for Quantum Computing, University of Waterloo\\
Waterloo, Ontario, Canada, N2L 3G1
}%

\date{\today}

\begin{abstract}
Decoupling the interactions in a spin network governed by a
pair-interaction Hamiltonian is a well-studied problem. Combinatorial
schemes for decoupling and for manipulating the couplings of
Hamiltonians have been developed which use selective pulses. In this
paper we consider an additional requirement on these pulse sequences:
as few {\em different} control operations as possible should be
used. This requirement is motivated by the fact that optimizing each
individual selective pulse will be expensive, i.\,e., it is desirable
to use as few different selective pulses as possible.  For an
arbitrary $d$-dimensional system we show that the ability to implement
only two control operations is sufficient to turn off the time
evolution. In case of a bipartite system with local control we show
that four different control operations are sufficient. Turning to
networks consisting of several $d$-dimensional nodes which are
governed by a pair-interaction Hamiltonian, we show that decoupling
can be achieved if one is able to control a number of different
control operations which is logarithmic in the number of nodes.
\end{abstract}

\pacs{{03.67.Lx, 03.65.Fd, 03.67.-a}}

\maketitle

%
%

\section{Introduction}

The problem of simulating pair-interaction Hamiltonians has been
studied intensively
\cite{JK:99,BCLLLPV:2002,Leung:2002,WRJB:2002,NBDCD:2002,DNBT:2002}.
It has been shown that starting from a given entangling bipartite
Hamiltonian any bipartite Hamiltonian can be simulated
\cite{BCLLLPV:2002}. This can be extended to networks of
qubits \cite{JK:99,Leung:2002,DNBT:2002} and even to networks of
higher dimensional systems \cite{WRJB:2002,NBDCD:2002}. The underlying
idea is to control the system by applying hard pulses to the
individual nodes of the network. This parallels the methods developed
for universal control of open quantum systems
\cite{VL:98,VKL:99,VLK:99,Zanardi:2001}. Decoupling of arbitrary
interactions is an important primitive for the simulation. This can be
achieved by using combinatorial constructions such as triples of
Hadamard matrices \cite{Leung:2002} and orthogonal arrays
\cite{SM:2001}. The control operations in these schemes are local and
have to be applied selectively to the nodes. In the
following we give a brief account of the general setting.
 
We assume that $H$ is a pair-interaction Hamiltonian of a system which
consists of $N$ coupled spin-$\frac{1}{2}$ particles. Let
$\sigma^{(i)}_{\alpha}$, where $\alpha \in \{ x, y, z \}$ and $i \in
\{1, \ldots, N\}$ denote the Pauli operators acting locally on qubit
$i$. We can write $H$ in the form
\begin{equation}\label{pairinteract}
H = \sum_{k=1}^N \sum_{\alpha}
r_\alpha^{(k)} \sigma_\alpha^{(k)} + \sum_{k,\ell=1}^N \sum_{\alpha,
\beta} J_{\alpha,\beta}^{k,\ell} \sigma_\alpha^{(k)} \otimes 
\sigma_\beta^{(\ell)},
\end{equation}
with indices $\alpha, \beta \in \{x, y, z\}$.  Hence the coupling
strength between the different qubits in this $N$-spin network is
given by $J_{\alpha,\beta}^{k,\ell}$. The problem at hand is to
decouple interactions in a Hamiltonian of the form
eq.~(\ref{pairinteract}), or more generally to simulate another
Hamiltonian $\tilde{H}$ by the given one. The framework for these
simulations is average Hamiltonian theory \cite{Slichter:90,EBW:87}.

A special case is given by a system which has $\sigma_z \sigma_z$
interactions only. Here selective $\sigma_x$ pulses applied to the
nodes are sufficient to turn off the interactions and hence to
simulate any desired Hamiltonian \cite{JK:99,Leung:2002}.

For general pair-interaction Hamiltonians the decoupling problem is
harder. Nevertheless, in \cite{SM:2001} it was shown how to achieve
decoupling using orthogonal arrays which are objects studied in
combinatorial theory. The approach taken in \cite{SM:2001} generalizes
schemes obtained from Hamadard matrices \cite{JK:99,Leung:2002}. All
these methods can be thought of as generalizations of well-known
techniques for decoupling and refocusing used in
nuclear-magnetic-resonance theory \cite{Slichter:90,EBW:87}. In
higher-dimensional systems decoupling can be achieved by means of
selective pulses which are derived from orthogonal arrays
\cite{SM:2001,WRJB:2002}. The selective pulses employed in these
schemes will be quite demanding in experimental realizations. Hence it
would be desirable to have as few different selective pulses as
possible in order to minimize the optimization overhead necessary to
implement each one of them. This motivates the question to search for
decoupling schemes in which the pulses can be arranged in such a way
that only few different pulses are used.

After a brief introduction into the framework we will consider schemes
for one $d$-dimensional node which will work with two different
pulses, schemes for a bipartite system of two $d$-dimensional nodes
which will work with four different pulses, and finally schemes for
networks of $n$ nodes which are $d$-dimensional. The latter schemes
make use of $O(\log n)$ different pulses.

%
%

\section{Decoupling Schemes}

The underlying model of the decoupling schemes described in this paper
is average Hamiltonian theory \cite{EBW:87}, the necessary parts of
which we briefly describe next. Assume that the Hamiltonian $H$ is of
the form (\ref{pairinteract}). We can apply the sequence
\begin{equation}\label{eqn1}
e^{-i\tau_n H} V_n \ldots e^{-i \tau_2 H} V_2 e^{-i \tau_1 H} V_1,
\end{equation}
with relative times $\tau_i\in \R$ and local unitaries $V_i$ for $i=1,
\ldots, n$. Note that the requirement $\prod_{i=1}^n V_i = \onemat$ is
necessary in order for average Hamiltonian theory to hold and to
determine the terms in the Magnus expansion \cite{Haeberlen:76} of the
piece-wise constant evolution given in eq.~(\ref{eqn1}). We can
rewrite (\ref{eqn1}) in the form
\[
(V_n \ldots V_1)^\dagger
e^{-i \tau_n H} 
\ldots 
 (V_2 V_1)^\dagger e^{-i \tau_2 H} (V_2 V_1)\;
V_1^\dagger e^{-i\tau_1 H} V_1,
\]
i.\,e., as a product in which all factors are given by Hamiltonians
$H_j = U_j H U_j^\dagger$, where $U_j := \prod_{k=1}^j V_k$ (ordered
from the right). Hence, the time evolution is divided into $n$
intervals in each of which we have a conjugated time evolution
$U_j^\dagger e^{-i \tau_k H} U_j$ with respect to the basis given by
$U_j$, i.\,e., this is the ``toggling-frame'' form of the sequence
\cite{EBW:87}.  We can still rewrite this using $U_j = \prod_{k=1}^j
V_k$ and obtain the new sequence
\begin{equation}\label{eqn2}
U_n^\dagger e^{-i\tau_n H} U_n U_{n-1}^\dagger \ldots 
U_2^\dagger e^{-i \tau_2 H} U_2 U_1^\dagger e^{-i \tau_1 H} U_1.
\end{equation}
This is the form of the pulse sequence we actually work with in the
following. The effective Hamiltonian corresponding to
eq.~(\ref{eqn2}), i.\,e., the first term in the Magnus expansion, is
given by $\sum_{i=1}^n U_i^\dagger H U_i$.  If the unitary operators
$U_i$ applied in the toggling-frame have the property that for each
pair $(k,l)$ of nodes the respective local operations applied to spins
$k$ and $l$ run through the elements of a unitary operator basis
${\cal B}$ \cite{Schwinger:60,Knill:96,KR:2002a} of the subsystem
given by $k$ and $l$, then we obtain that the time evolution of the
system is stopped.  Indeed, on the subsystem we then obtain the
average Hamiltonian \cite{WRJB:2002}
\[
\frac{1}{|{\cal B}|} \sum_{U \in {\cal B}}
U^\dagger H U = \frac{1}{d} {\rm tr}(H) \onemat_d,
\]
which is zero since ${\rm tr}(H)=0$. Unitary operator bases ${\cal B}$
exist in any dimension. Indeed, let $d\geq 2$ be the dimension of the
system, we explicitly define a basis for the vector space of
$\C^{d\times d}$ matrices: let $\omega_d = \exp(2 \pi i/d)$, $\sigma_x
= \sum_{i=0}^{d-1} \ket{i}\bra{i+1}$, $\sigma_z = \sum_{i=0}^{d-1}
\omega_d^i \ket{i}\bra{i}$, where all indices are computed modulo
$d$. Then ${\cal B} = \{ \sigma_x^i \sigma_z^j : i,j=0,\ldots, d-1\}$
is an operator basis which will be referred to as the Pauli basis.

We continue with an observation on the sequence (\ref{eqn2}): while
the matrices $U_i$ have to run through the list of all elements of the
operator basis ${\cal B}$, the list of quotients $U_{i+1} U_i^\dagger$
can be considerably smaller. The following example shows that for a
system consisting of two $d$ dimensional subsystems it is enough to be
able to apply four different operators in order to switch off the time
evolution.

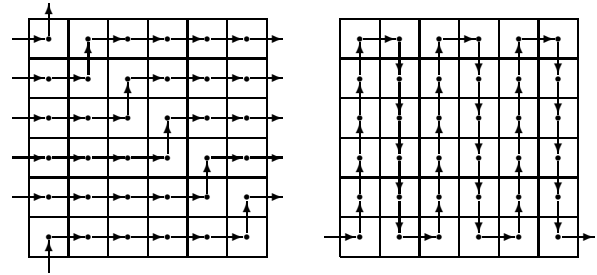
\begin{figure}%
\unitlength1.5pt
\begin{tabular}{c@{\qquad\quad}c}
\begin{picture}(60,60)%
\multiput(0,0)(10,0){7}{\line(0,1){60}}
\multiput(0,0)(0,10){7}{\line(1,0){60}}
\multiput(5,5)(10,0){6}{%
\multiput(0,0)(0,10){6}{\circle*{1}}}
\put(5,-4){\vector(0,10){8}}
\multiput(6,5)(10,0){5}{\vector(1,0){8}}
\multiput(6,15)(10,0){4}{\vector(1,0){8}}
\multiput(6,25)(10,0){3}{\vector(1,0){8}}
\multiput(6,35)(10,0){2}{\vector(1,0){8}}
\multiput(6,45)(10,0){1}{\vector(1,0){8}}
\multiput(56,15)(10,0){1}{\vector(1,0){8}}
\multiput(46,25)(10,0){2}{\vector(1,0){8}}
\multiput(36,35)(10,0){3}{\vector(1,0){8}}
\multiput(26,45)(10,0){4}{\vector(1,0){8}}
\multiput(16,55)(10,0){5}{\vector(1,0){8}}
\multiput(-4,15)(0,10){5}{\vector(1,0){8}}
\put(55,6){\vector(0,1){8}}
\put(45,16){\vector(0,1){8}}
\put(35,26){\vector(0,1){8}}
\put(25,36){\vector(0,1){8}}
\put(15,46){\vector(0,1){8}}
\put(5,56){\vector(0,1){8}}
\end{picture}
&\begin{picture}(60,60)%
\multiput(0,0)(10,0){7}{\line(0,1){60}}
\multiput(0,0)(0,10){7}{\line(1,0){60}}
\multiput(5,5)(10,0){6}{%
\multiput(0,0)(0,10){6}{\circle*{1}}}
\put(-4,5){\vector(1,0){8}}
\multiput(5,6)(0,10){5}{\vector(0,1){8}}
\multiput(25,6)(0,10){5}{\vector(0,1){8}}
\multiput(45,6)(0,10){5}{\vector(0,1){8}}
\multiput(15,54)(0,-10){5}{\vector(0,-1){8}}
\multiput(35,54)(0,-10){5}{\vector(0,-1){8}}
\multiput(55,54)(0,-10){5}{\vector(0,-1){8}}
\put(6,55){\vector(1,0){8}}
\put(16,5){\vector(1,0){8}}
\put(26,55){\vector(1,0){8}}
\put(36,5){\vector(1,0){8}}
\put(46,55){\vector(1,0){8}}
\put(56,5){\vector(1,0){8}}
\end{picture}
\end{tabular}
\caption{\label{hamiltonCycle} Decoupling of a Hamiltonian on $\C^d$
can be achieved using two different pulses (the case $d=6$ is shown).
Pulses correspond to elements in $\Z_d^2$ which is generated up to
phases by $\sigma_x$ and $\sigma_z$. The left
square shows the sequence in which the pulses have to be applied. In
each square a horizontal move $\rightarrow$ corresponds to
multiplication with $\sigma_x^{(i)}$ and a vertical move $\uparrow$ to
multiplication with $\sigma_z^{(i)}$ for $i=1,2$. An alternative
sequence is given in the right square. Decoupling of a bipartite
Hamiltonian on $\C^d \otimes \C^d$ can be achieved using four
different pulses. Here pulses correspond to elements in $\Z_d^4 =
\Z_d^2 \times \Z_d^2$ generated by $\sigma_x^{(i)}$,
$\sigma_z^{(i)}$ for $i=1,2$. A cyclic sequence can be obtained by
performing one step in the first copy of $\Z_d^2$ followed by one full
cycle in the second copy and so on.}
\end{figure}

{\it Example.}  In Figure \ref{hamiltonCycle} we have shown how to
label the elements of the Pauli basis for a $d$ dimensional system in
such a way that we reach all elements by just multiplying one
generator at a time. It is also shown how the elements of an operator
basis for a bipartite system can be arranged. This corresponds to an
enumeration of the elements of $\Z_d^4 = \langle S \rangle/(\langle S
\rangle \cap \C^*)$, where $S := \{ \sigma_x^{(1)},
\sigma_z^{(1)},\sigma_x^{(2)}, \sigma_z^{(2)}\}$. This enumeration can
be cast in terms of Hamilton cycles in Cayley graphs (for the basic
notions of graph theory we refer to \cite{GR:2001}). Let $G$ be a
group and let $S$ be a set of generators of $G$. Then the Cayley graph
$\Gamma(G, S) = (V,E)$ is the directed graph with vertices $V=\{v_g
: g \in G\}$ labeled by the group elements. The edges $E$ are as
follows: there is a directed edge from $v_g$ to $v_h$ if there exists
$s \in S$ such that $h = sg$. Each Cayley graph is regular of degree
$|S|$ and its structure depends on the specific choice of the
generating set $S$.

A tour through the vertices of a graph which visits each vertex
precisely once is called a Hamilton path. In case there is an edge
between the end node and the start node we can close the cycle and
obtain a Hamilton cycle. The sequences shown in Figure
\ref{hamiltonCycle} define Hamilton cycles in the Cayley graph
$\Gamma(\Z_d^4, S)$.

While for graphs like the Cayley graph of $\Z_d^4$ Hamilton cycles can
easily constructed directly, the general question whether Hamilton
cycles exist in arbitrary Cayley graphs is open \cite{CG:96}. In
general the problem to decide whether a given graph contains a
Hamilton cycle is a difficult problem and of interest in computer
science and optimization. It is one of the classical NP complete
problems \cite{PS:82}, i.\,e., it is believed that no polynomial time
exists for this problem.

For the rest of the paper we use the fact that there are Hamilton
cycles in the Cayley graphs $\Gamma(G,S)$ for the abelian groups
$G=\Z_d^{2n}$ and the set $S$ of generators is
given by the $2n$ coordinate vectors. These can be obtained by
generalizing the construction given in Fig.~\ref{hamiltonCycle}.

%
%

\section{Orthogonal Arrays}

In the design of statistical experiments \cite{BJL:99I} which depend
on several factors one is often forced to work with an incomplete
subset of the set of all possible combinations of factors. Orthogonal
arrays provide a way to plan such experiments systematically. An
orthogonal array is a (in general non-square) matrix with entries from
a finite set $S$ of symbols. The properties of an array are determined
by a set of characteristic parameters for which the notation
$OA(N,n,s,t)$ is used \cite{HSS:99}. An $n \times N$ matrix $A$ with
entries from the set $S$ is an orthogonal array with $s := |S|$ levels
and strength $t$ iff every $t \times N$ subarray of $A$ contains each
possible $t$-tuple of elements in $|S|$ precisely once as a column of
$A$. The shorthand notation for this is that $A$ is an $OA(N, n, s,
t)$.

The rows of this matrix are also called {\em factors} and correspond
to nodes of the network. The columns are also called {\em runs} and
correspond to the time slots of the scheme, i.\,e., to the pulses to
be applied. The most important parameter is the {\em strength} $t$ of
the array. In the context of simulation of pair-interaction
Hamiltonians the strength has to be $t=2$.

In \cite{SM:2001} it has been shown that given an
orthogonal array $OA(N, n, 4, 2)$, an arbitrary and possibly unknown
Hamiltonian $H$, which describes the pair-interactions in an $n$ qubit
network, can be decoupled using $N$ pulses. 

We now show that in this case the pulse sequence can be arranged in
such a way that only a minimal number of different pulses have to be
applied. First, we briefly recall some basic facts about
error correcting codes \cite{MS:77} since they will feature in the
subsequent constructions of orthogonal arrays. A linear code over the
finite field $\F_q$ is a $k$-dimensional subspace of the vector space
$\F_q^n$. The metric on the space $\F_q^n$ is called the Hamming
weight which for $x = (x_1, \ldots, x_n) \in \F_q^n$ is defined by
${\rm wt}(x) := |\{ i\in \{1, \ldots, n\} : x_i \not=0\}|$. The
minimum distance of a linear code $C$ is defined by $d = d_{\rm min}
:= \min{\{\rm wt}(c) : c \in C\}$. As a shorthand we abbreviate this
by saying that $C$ is an $[n, k, d]_q$ code. We need one more
definition which is the dual code $C^\perp$ of $C$ defined by $C^\perp
:= \{ x \in \F_q^n : x \cdot y = 0 \; \mbox{for all} \; y \in
\F_q^n\}$.

The following theorem \cite[Theorem 4.6]{HSS:99} establishes a
connection between orthogonal arrays and error correcting codes. In
fact this is one of the most prolific construction for orthogonal
arrays known.

\begin{theorem}\label{OAcodes}
Let $C$ be a linear $[n, k, d]_q$ code over $\F_q$. Let $d^\perp$ be
the minimum distance of $C^\perp$. Arrange the codewords of $C$ into
the columns of a matrix $A \in \F_q^{n\times q^k}$. Then $A$ is an
$OA(q^k, n, q, d^\perp-1)$.
\end{theorem}

Each column of the orthogonal array corresponds to one of the pulses
$U_i$ in eq.~(\ref{eqn2}) and we want to keep the set $\{U_{i+1}
U_i^\dagger\}$ as small as possible. We are going to show next that
for the arrays constructed from Theorem \ref{OAcodes} this is always
possible.

%
%

\section{Decoupling using Gray Codes}

We illustrate Theorem \ref{OAcodes} by a scheme for a small network
consisting of $5$ qubits. The finite field needed for this case is the
field $\F_4 =\{0, 1, \omega, \overline{\omega}\}$ of four elements in
which the relation $1+\omega+\overline{\omega} = 0$ holds. The
following is a generator matrix for the quadratic residue code
\cite{MS:77} $C$ over $\F_4$ with parameters $[5,2,4]$.
\[
\left[
\begin{array}{ccccc}
1 & 0 & 1 & \overline{\omega} & \overline{\omega} \\
0 & 1 & \overline{\omega} & \overline{\omega} & 1 
\end{array}
\right]
\]

The dual code $C^\perp$ of this code again is defined over $\F_4$ and
has parameters $[5,3,3]$. 
By taking all $4^2 = 16$ code-words in $C$ as the columns of a matrix
we obtain an $OA(16,5,4,2)$. We now address the question how to
arrange the columns of this $4\times 16$ matrix. First, recall the
Gray codes \cite{Gilbert:58,Savage:97} are Hamilton cycles for the
groups $\F_2^n$ where the generator sets are the coordinate
functions. For $\F_2^4$ a particular Gray code is given by the
following sequence of binary strings:
\begin{equation}\label{gray}
\left[
\begin{array}{cccccccccccccccc}
0 & 0 & 0 & 0 & 0 & 0 & 0 & 0 & 1 & 1 & 1 & 1 & 1 & 1 & 1 & 1 \\ 
0 & 0 & 0 & 0 & 1 & 1 & 1 & 1 & 1 & 1 & 1 & 1 & 0 & 0 & 0 & 0 \\
0 & 0 & 1 & 1 & 1 & 1 & 0 & 0 & 0 & 0 & 1 & 1 & 1 & 1 & 0 & 0 \\
0 & 1 & 1 & 0 & 0 & 1 & 1 & 0 & 0 & 1 & 1 & 0 & 0 & 1 & 1 & 0
\end{array}
\right].
\end{equation}
Observe that from column $c_i$ to $c_{i+1}$ exactly one position is
flipped. Overall we obtain a cyclic sequence which runs through all
$16$ elements of $\F_2^4$ exactly once. The order defined by the Gray
code also defines an order of the vectors in $\F_4^2$. To obtain the
corresponding elements in $\F_4^2$ we make the identification $(0,0)
\mapsto 0$, $(0,1) \mapsto 1$, $(1,0) \mapsto \omega$, and $(1,1)
\mapsto \overline{\omega}$. Using this identification we can now map
all elements of $\F_4^2$ to elements of $\F_4^5$ using the linear code
$C$. 
We obtain the following scheme for a system of five qubits:
\[
\left[
\begin{array}{cccccccccccccccc}
0 & 0 & 0 & 0 & 1 & 1 & 1 & 1 & \overline{\omega} & \overline{\omega} & \overline{\omega} & \overline{\omega} & \omega & \omega & \omega & \omega \\
0 & 1 & \overline{\omega} & \omega & \omega & \overline{\omega} & 1 & 0 & 0 & 1 & \overline{\omega} & \omega & \omega & \overline{\omega} & 1 & 0 \\
0 & \overline{\omega} & \omega & 1 & 0 & \overline{\omega} & \omega & 1 & \overline{\omega} & 0 & 1 & \omega & \overline{\omega} & 0 & 1 & \omega \\
0 & \overline{\omega} & \omega & 1 & \omega & 1 & 0 & \overline{\omega} & \omega & 1 & 0 & \overline{\omega} & 0 & \overline{\omega} & \omega & 1 \\
0 & 1 & \overline{\omega} & \omega & 1 & 0 & \omega & \overline{\omega} & \omega & \overline{\omega} & 1 & 0 & \overline{\omega} & \omega & 0 & 1 
\end{array}
\right]
\]
Hence there are four basic pulses $\pi_1, \ldots, \pi_4$ corresponding
to those columns in (\ref{gray}) which are elementary vectors $e_1,
\ldots, e_4$, i.\,e., columns $2$, $4$, $8$, and $16$. The
corresponding operators are $\pi_1 = \onemat_2 \otimes \sigma_1
\otimes \sigma_3 \otimes \sigma_3 \otimes \sigma_1$, $\pi_2 =
\onemat_2 \otimes \sigma_2 \otimes \sigma_1 \otimes \sigma_1 \otimes
\sigma_2$, $\pi_3 = \sigma_1 \otimes \onemat_2 \otimes \sigma_1
\otimes \sigma_3 \otimes \sigma_3$, $\pi_4 = \sigma_2 \otimes
\onemat_2 \otimes \sigma_2 \otimes \sigma_1 \otimes \sigma_1$.  The
order in which they have to be applied can be read off from the
transitions in (\ref{gray}), i.\,e., we obtain the following pulse
sequence:
\[
\pi_1, \pi_2, \pi_1, \pi_3, \pi_1, \pi_2, \pi_1, \pi_4,
\pi_1, \pi_2, \pi_1, \pi_3, \pi_1, \pi_2, \pi_1, \pi_4.
\]

We now turn to the question how to generalize this to more general
networks and show that pulse sequences obtained from linear codes have
the property that the number of different pulses grows logarithmically
with the number of spins.  In order to apply Theorem \ref{OAcodes} we
have to find linear codes over $\F_4$ for which the minimum distance
of the dual code is at least $3$.

Let $q$ be a prime power and let $m \in \N$. The Hamming code
$H_{q,m}$ of length $n:=(q^m{-}1)/(q{-}1)$ is a single-error
correcting linear code with parameters $[n,n-m,3]_q$. The
corresponding dual code $H^\perp_{q,m}$ has parameters
$[n,m,q^{m-1}]$. By specializing $q=4$ and by using Theorem
\ref{OAcodes} for $H^\perp_{4,m}$ we therefore obtain orthogonal
arrays with parameters $OA(N, n, 4, 2)$, where $n=(4^m{-}1)/3$ and $N
= 4^{n-m}$ for any choice of $m \in \N$.

The procedure to obtain a pulse sequence for a network of $n$
spin-$\frac{1}{2}$ particles, where $n_0$ is an arbitrary natural
number, i.\,e., not necessarily of the form $n=(4^m-1)/3$ is as
follows: first let $m\in \N$ be the unique integer such that $n_0 \leq
\frac{4^m-1}{3} \leq 4n_0$. Then construct the orthogonal array with
parameters $OA(4^m, (4^m-1)/3,4,2)$. The columns of this orthogonal
array are codewords of $H_{4,m}^\perp \subseteq \F_4^{(4^m-1)/3}$.  We
will use only the first $n_0$ rows of this $\frac{4^m-1}{3}-m \times
\frac{4^m-1}{3}$ array. Since $H_{4,m}^\perp \cong \F_4^m$, we can
find the desired Hamilton cycle by choosing a Gray code on
$\F_2^{2m}$, i.\,e., we can decouple using $2m$ different pulses.

%
%

{\it Higher Dimensional Systems.---} So far we have only considered
spin-$\frac{1}{2}$ particles, i.\,e., networks consisting of
qubits. Methods exist to decouple Hamiltonians also in case the
dimension of the individual nodes is greater than two
\cite{WRJB:2002,NBDCD:2002}. Basically, for one node of dimension
$d\geq 2$ the requirement to switch off the time evolution of this
node is to apply all $d^2$ elements of a unitary operator basis
$\{U_{i,j} \, : \, i,j=0,\ldots,d-1\}$. For any pair of nodes we have
to apply all $d^4$ elements of a tensor product basis $\{U_{i,j}
\otimes U_{k,\ell} \, : \, i,j,k,\ell=0,\ldots,d-1\}$. These pulses
can be arranged like in the previous schemes using orthogonal
arrays. By combining the construction of orthogonal arrays from
Hamming codes over the alphabet $\F_{2^{2\alpha}}$, where $\alpha\in \N$,
together with the Gray codes for $\F_{2^{2m}}$, which can be
constructed as in Figure \ref{hamiltonCycle}, we obtain schemes for
networks of $n=(2^{2m}-1)/3$ nodes (each of dimension $2^\alpha$) which use
only ${4 \log n}$ different pulses.

{\it Conclusions and Discussion.---} We have shown that pulse
sequences for decoupling in networks of qubits and higher-dimensional
systems can be arranged in such a way that the number of different
pulses needed is small. Using orthogonal arrays derived from Hamming
codes over finite fields it has been shown that the number of
different operations can be chosen to be of the order $O(\log n)$,
where $n$ is the number of nodes in the network. The method presented
in this paper is based on Hamilton cycles in the Cayley graph of the
group defined by the columns of an orthogonal array. The related graph
theoretical concept of Eulerian cycles has been used recently for the
problem of designing robust schemes for decoherence control
\cite{VK:2003,Viola:2004}. Since in this case the Hamiltonian of the
system can be arbitrary, i.\,e., not necessarily of pair-interaction
type, the task to decouple from an environment leads to a different
control scenario. It would be of interest to determine whether the
methods described in this paper can also be used to derive efficient
schemes for robust Eulerian dynamical coupling.

The author thanks E.~Knill, R.~Laflamme, and P.~Zanardi for valuable
discussions. This work was supported in part by NSA and ARDA under the
ARDA Quantum Computing Program. The author also acknowledges support
by CFI, ORDCF, and MITACS.

\end{document}